\documentclass[twocolumn,prl]{revtex4}%
\pdfoutput=1

\usepackage{epsfig}
\usepackage{graphics}
\begin{document}
\title{Measurement of Quantum Geometry Using Laser Interferometry}
\begin{abstract}
New quantum degrees of freedom of space-time, originating at the Planck scale,  could create a coherent indeterminacy and noise in the transverse position of massive bodies on macroscopic scales.  An experiment  is under development at Fermilab designed to detect or rule out a transverse position noise with Planck spectral density, using correlated signals from an adjacent pair of  Michelson interferometers. A detection would open an  experimental window on quantum space-time.

\end{abstract}

\maketitle
 Quantum effects of space-time are predicted to originate at  the 
 Planck scale, $ct_P\equiv \sqrt{\hbar G/c^3}= 1.616\times 10^{-35}$m.   In standard quantum field theory,   their effects are strongly suppressed at  experimentally accessible energies, so space-time is predicted to behave almost classically, for practical purposes, in  particle experiments.
However,  new 
quantum effects of geometry  originating at the Planck scale--- from geometrical degrees of freedom not included in standard field theory--- may have effects on macroscopic scales that could be measured by laser interferometers. 

The possibility of new quantum-geometrical degrees of freedom is suggested from several theoretical directions. Quantum physics is experimentally proven to violate the principle of locality on which classical space-time is based. Gravitational theory suggests that quantum states of space-time systems do not respect locality of the kind assumed by quantum field theory, and  suggests that space-time and gravity  are approximate statistical behaviors of a quantum system with a holographic information content, far less than that predicted by quantum field theory.\cite{Jacobson:1995ab,Verlinde:2010hp}  

Quantum geometry could arise in Planck scale physics, but still produce a detectable displacement in a macroscopic experiment.\cite{Hogan:2012ib}   A typical uncertainty in wave mechanics, if  information about transverse position is transmitted nonlocally with a bandwidth limit, is the scale familiar from diffraction-limited imaging: the geometric mean of inverse bandwidth and  apparatus size. For separations on a laboratory scale, a Planck scale frequency limit   leads to a transverse uncertainty in position on the order of attometers.  Displacements of massive bodies of this order are detectable using laser interferometry.

No fundamental theory of quantum geometry exists, but a consistent effective theory, based on general properties of quantum mechanics and covariance,  can be used to precisely predict a phenomenology on macroscopic scales.  In particular, the theory precisely relates the number of geometrical position eigenstates to the amplitude of indeterminacy in  transverse position at separation $L$, so it can be related to the holographic density of states predicted from gravitational theory.  This hypothesis leads to an exact  prediction for the variance in transverse position with no free parameters,\cite{Hogan:2012ne} 
\begin{equation}
\langle x_\perp^2 \rangle= L ct_P/ \sqrt{4\pi} .
\end{equation}
Planckian  indeterminacy leads to a new form  of noise in position with this displacement, on a timescale $L/c$.  This form of indeterminacy would have escaped detection to date, and indeed is overwhelmed by standard quantum indeterminacy on the mass scale of elementary particles.  However, it is detectable as a new source of quantum-geometrical noise in an interferometer that coherently measures the positions of massive bodies in two directions over a macroscopic volume.\cite{Hogan:2010zs,Hogan:2012ib}  

The Fermilab Holometer is an experiment (E-990), funded largely through an Early Career Award to Aaron Chou at Fermilab, designed to detect or rule out quantum-geometrical noise with these properties.\cite{holowebsite}
Much of the  technology  has been developed by LIGO and other projects to measure displacements due to gravitational radiation.  The quantum-geometrical measurement however calls for application of the  technology in a new  experimental design.  Measurements can be made at relatively high (MHz) frequencies, where environmental and gravitational noise sources are smaller, both shrinking and simplifying  the layout. The experiment is designed to measure the specific and peculiarly quantum-mechanical signatures of the effect, such as nonlocal coherence and transverse nature of the indeterminacy,  the frequency cross spectrum, and time-domain cross correlation function. It is anticipated that the experiment will be complete, and either detect or rule out this  form of Planckian noise, within about two years.

If the noise is found  not to exist, only a modest followup effort may be motivated to pursue the limits somewhat past the Planck scale for a conclusive result. If it is found, a significantly expanded experimental program can be pursued to obtain high precision results and map out the spatiotemporal properties of quantum geometry.

\end{document}